\documentclass[twocolumn,secnumarabic,amssymb, nobibnotes, aps, prd]{revtex4-2}

\usepackage{amsmath}
\usepackage{graphicx}
\usepackage{amsfonts}
\usepackage{amssymb}
\usepackage{fancybox}
\usepackage{fancyhdr}
\usepackage{bm}
\usepackage{color}
\usepackage{titlesec}

\newcommand{\be}{\begin{equation}}
\newcommand{\ee}{\end{equation}}
\newcommand{\beq}{\begin{eqnarray}}
\newcommand{\eeq}{\end{eqnarray}}
\newcommand{\n}{\nonumber}

\newcommand{\sk}{\smallskip \noindent}

\newcommand{\wt}{\widetilde{\nabla}}
\newcommand{\td}{\tilde{\delta}_\perp}

\begin{document}

\title{Jacobi equations of geodetic brane gravity }

\author{Riccardo Capovilla$^{\ddagger}$, 
Giovany Cruz${}^{\dagger}$, and Efra\'\i n Rojas}%
\email{capo@fis.cinvestav.mx$^\ddagger$, gcruz@fis.cinvestav.mx$^\dagger$, efrojas@uv.mx$^*$}
\affiliation{Departamento de F\'{\i}sica, Cinvestav-IPN, 
Av. Instituto Polit\'ecnico Nacional 2508,
Col. San Pedro Zacatenco, 07360, Gustavo A. Madero, Ciudad 
de M\'exico, M\'exico.$^{\ddagger \dagger}$
\\
Facultad de Física, Universidad Veracruzana, Campus Sur, 
Paseo No. 112, Desarrollo Habitacional, Nuevo Xalapa, 91097, 
Xalapa-Enr\'\i quez, Veracruz, M\'exico.$^*$}
\date{April 4, 2022}%
\begin{abstract}
We consider brane gravity as 
described by the Regge-Teitelboim geometric model, in any 
co-dimension. In brane gravity our spacetime is modeled as 
the time-like world volume spanned by a space-like brane in 
its evolution,  seen as a manifold embedded in an ambient 
background Minkowski spacetime of higher dimension. Although 
the equations of motion of the model are well known, apparently 
their linearization has not been considered before. Using a 
direct approach, we linearize the equations of motion about a solution,
obtaining the Jacobi equations of the Regge-Teitelboim model.
They take a formidable aspect. Some of their features are 
commented upon.  By identifying the Jacobi equations, we derive 
an explicit expression for the Morse {\it index} of the model. 
To be concrete, we apply the Jacobi equations to the study of 
the stability of a  four-dimensional  Schwarzschild spacetime 
embedded in a six-dimensional Minkowski spacetime. We find that 
it is unstable under small linear deformations.
\end{abstract}

\maketitle

\section{Introduction}

In the 70's,  T. Regge and C. Teitelboim  (RT) considered 
a geometric model for our spacetime as the world volume of 
a three-dimensional brane evolving  geodesically in a fixed 
higher-dimensional background Minkowski spacetime. For their 
motivation, the title of their proceedings contribution, 
{\it Gravity \`a la string: a progress report}, says it all ~\cite{RTmodel}. The action they considered  in their geometric 
model is identical to the Einstein-Hilbert action of general 
relativity. The crucial difference are the field variables. 
Rather than the spacetime metric as in general relativity, 
in the RT model the field variables  are the embedding 
functions of the world volume, so that the world volume metric 
becomes  a composite field variable. 

The equations of motion 
of the RT model are of second order in derivatives, and weaker 
than the Einstein equations. The feature of geometric models with equations of motion of second order is
shared by a larger class of geometric models, to which the RT model belongs, called Lovelock branes \cite{goon2011new, cruz2013born}. 
All solutions of the Einstein 
equations are also solutions of the RT model, but the solution 
space of the latter is larger \cite{maia1986,pavsic1986,tapia1989}. 
The extra part can  be interpreted as ``geometrical dark matter" \cite{davidson2001cold}, in competition with present efforts  
to describe dark matter/energy that add  exotic terms to the 
energy-momentum tensor \cite{darkenergydynamics,darkmatter1},
or modifications of the geometric part, like  $f(R)$ theories, 
see {\it e.g.} \cite{fR1,fR2,fR3}. In this context, we note that 
the addition of a world volume cosmological constant is 
equivalent, from a brane point of view, to the addition of a 
Dirac-Nambu-Goto (DNG) term to the action.

The pioneering work of Regge and Teitelboim has received 
relatively recently renewed interest in the context of brane-world scenarios \cite{arkani1998hierarchy, randall1999large, maartens2010}, and in 
particular through the studies of Davidson and collaborators, 
that introduced the suggestive term ``Geodetic Brane Gravity" , 
see {\it e.g.}~\cite{davidson1998quantum, karasik2003geodetic}. In 
addition, one expects that the RT action will emerge
as an effective action in any geometrical treatment of 
branes that takes into account their physical thickness. At the same 
order of magnitude, one finds possible geometric models quadratic 
in the extrinsic curvature tensor, known as ``rigidity terms" ~\cite{polyakov1986fine, carter1995curvature}.

It is important to note that, at a basic geometric level, in 
order to ensure the local existence of an embedding framework, 
at most $N = n(n+1)/2$ dimensions are needed for the ambient 
spacetime background. For $n=4$, at most 10 dimensions are 
needed.  In addition, it is known that if the world volume 
metric admits Killing vectors, that number can be reduced~\cite{janet1926possibilite,cartan1927possibilite,friedman1961local, rosen1965embedding}. 
In particular, we remark that not every spacetime solution 
of the Einstein equations can be embedded as a hypersurface. 
For example the embedding of the Schwarzschild solution requires 
at least co-dimension two \cite{paston2012embeddings}. This 
provides one important motivation to consider arbitrary co-dimension, 
even though it does complicate things. Needless 
to say,  this is relevant about the stability of this type of geometric 
configurations, since it is also necessary to analyse the 
conditions to ensure or not its stability. 
In particular, higher co-dimension implies 
the necessity to use geometrical structures that take into 
account the rotational freedom in the normal fields to the 
world volume. In our opinion, this issue was 
overlooked in many contributions on the RT model, and does not 
appear to have been addressed before.

The aim of this note is to derive the Jacobi equations for 
the RT geometric model. For a relativistic particle that 
satisfies the geodesic equation, the Jacobi equations are 
simply the geodesic deviation equation, that quantifies the 
behaviour of two nearby particle in a curved background spacetime \cite{wald2010general}. In the higher dimensional brane generalization, 
one can adopt the same point of view. More in general, the Jacobi equations provide an essential tool for the study of the stability 
of a solution of the equations of motion. 
The derivation of the Jacobi equations can be accomplished with a direct approach, that consists in the linearization of the equations 
of motion, and this the one adopted in this paper. We consider also the possible presence of matter fields that live on the brane.
 An equivalent and alternative avenue would be a variational approach, that considers the second variation of 
the action when the equations of motion are satisfied, {\it i.e.}  
on-shell. In both cases, the Jacobi equations can be derived 
as Euler-Lagrange equations of a suitable action principle 
that is a cousin of the Jacobi's accessory variational 
principle for systems with a finite number of degrees of freedom, see {\it e.g.}  \cite{kot2014first}. For the geometric model under consideration, this is accomplished by the identification
of interesting conserved geometrical structures.
The accessory action is proportional to the index 
of the geometrical model under consideration. The index is a crucial tool in the study of the geometry in the large,
using Morse theory, and it has proven to be a valuable tool in geometric stability problems, such as minimal surfaces, see {\it e.g.}
\cite{fomenko2005elements, colding2011course}. In addition, the index ought to  provide an interesting 
avenue towards a path integral quantization  of the model, 
since the accessory action is quadratic in the fields. 

Once established the general form of the Jacobi equations, 
we focus on a specific solution of the equations of motion,
namely a four dimensional Schwarzschild geometry embedded 
in a six dimensional Minkowski spacetime. Exploiting the
symmetries of the solution, we derive a set of equations 
that determine the quasi-normal modes of the system.
Following the guideline developed in~\cite{horowitz2000quasinormal},  a numerical calculation 
allows us to draw some conclusions 
about the linear stability of the geometry. 
Indeed, we find 
signals of instability for this configuration in absence of 
matter. Although this is a widely accepted feature in black hole 
theories in higher dimensions, it seems likely that stable 
solutions can be found  in higher-dimensional embedded black 
hole geometries~\cite{chamblin2000brane}.

This paper is organized as follows. In Sec. \ref{sec2} , a 
brief description of the geometry and  notation used is given. 
In Sec. \ref{sec3}, we introduce  the RT geometrical  model. 
In Sec. \ref{sec4}, we derive an explicit covariant form for 
the Jacobi equations of the model, via a direct linearization 
of the  equations of motion. We also derive the index of the 
model. In Sec. \ref{sec5}, we study the stability of a  four-dimensional Schwarzschild spacetime  embedded in a six-dimensional 
Minkowski spacetime background, using the results previously 
derived for the general case. Sec. \ref{sec6} provides a brief 
discussion. Some technical details are presented in two
Appendices.

\section{Brane geometry in higher co-dimension}
\label{sec2}

Consider a $p+1$ dimensional manifold $m$ that represents 
the world volume of a spacelike brane  $\Sigma$. $m$ is 
embedded in a $n+1$ dimensional flat Minkowski spacetime 
background {$\lbrace\mathcal{M}$,$\eta\rbrace$} with metric $\eta_{\mu\nu}=\text{diag}(-1,1,...,1)$
($\mu,\nu = 0,1,2, \ldots,n$). The world volume is described by the embedding functions $X^{\mu}(u^a)$ where $u^a$ are  local coordinates for $m$ ($a,b = 0,1,2, \ldots p$). The tangent vectors to $m$ are given by $X^{\mu}_a:=\partial X^{\mu}/\partial u^a$. The inner product of the
tangent  vectors gives the induced metric on $m$, $g_{ab}:=\eta_{\mu\nu}X^{\mu}_aX^{\nu}_b=X_a\cdot X_b$. 
Here and henceforth a dot  denotes 
inner product using the background Minkowski metric. 
 By $g^{ab}$ we denote the inverse of $g_{ab}$, and by $g$ its 
determinant. The world volume $m$ is assumed to be time-like so $g < 0$.
The normal vectors to $m$ are represented by $n^{\mu\,i}$, ($i,j=1,2,...,n-p$), and $n-p$ is the co-dimension of $m$. 
The normal vectors are orthogonal to the tangent vectors, $n^i\cdot X_a=0$,  and orthonormal between themselves, $n_i\cdot n_j=\delta_{ij}$. 
These expressions define the normal vectors up to a sign and a local $O(n-p)$ rotation. This gauge freedom requires the introduction
of a suitable gauge field, known as twist potential, given by
$\omega_a{}^{ij} = n^i \cdot \partial_a n^j$, see {\it e.g.} \cite{capovillageometry}.
We also introduce 
the extrinsic curvature of  $m$ with  $K_{ab}{}^i = - n^i \cdot \nabla_a 
X_b$, and the mean extrinsic curvature as its trace, $K^i = g^{ab} 
K_{ab}{}^i$.  Here $\nabla_a$ denotes the torsion-less 
world volume covariant 
derivative, compatible with the induced metric, $\nabla_a g_{bc}= 0$.
In addition, we will use the gauge covariant derivative given by 
 $\wt_a = \nabla_a - \omega_a$. The Riemann tensor of $\nabla_a$
 is denoted by $R_{abc}{}^d$, with Ricci tensor $R_{ab} = R_{acb}{}^c$,  scalar curvature
 $R = g^{ab} R_{ab}$, and Einstein tensor $G_{ab} = R_{ab} - (1/2) g_{ab} R $. We follow the conventions of  \cite{wald2010general}.

The intrinsic and extrinsic geometries of the world volume $m$ are related  by the Gauss-Codazzi-Mainardi  equations
\begin{subequations}
\label{whole}
\begin{equation}
R_{abcd} = K_{ac}{}^iK_{bd\,i}-K_{ab}{}^{i}K_{cd\,i},
\label{gcm1}
\end{equation}
\begin{equation}
0 = \wt_a K_{bc}{}^i -\wt_b K_{ac}{}^i.
\label{gcm2}
\end{equation}
\end{subequations}
For completeness, it should be mentioned that, for co-dimension
higher that one, there are additional equations that involve the curvature 
of the twist potential, see {\it e.g.} \cite{capovillageometry}, but they will not be 
used in this paper.

We note also the contractions with the covariant metric,
\begin{subequations}
\begin{equation}
R_{ab}=K^i K_{ab\,i}-K_{ac\,i} K_{b}{}^{c\,i},
\label{gcm3}
\end{equation}
\begin{equation}
R = K^i K_i - K^{ab\,i} K_{ab\,i},
\label{gcm4}
\end{equation}
\begin{equation}
0 = \wt_a K^i - \wt^b K_{ab}{}^i\,.
\end{equation}
\label{whole1}
\end{subequations}
Additionally, for any tensor $S_{ab}$, $S_{[ab]}$ indicates
anti-symmetrization under the convention $S_{[ab]} = (S_{ab} 
- S_{ba})/2$. Similarly, $S_{(ab)}$ indicates symmetrization 
according to $S_{(ab)} = (S_{ab} + S_{ba})/2$.

\section{Regge-Teitelboim  model}
\label{sec3}

The RT model is the integral over the trajectory 
of a $p$-dimensional brane $\Sigma$, that depends on
the scalar curvature $R$ of the world volume $m$ obtained from the
world volume metric $g_{ab}$
\citep{RTmodel},
\begin{equation}
S_{\text{\tiny RT}}[X,\varphi]= \int_m 
\sqrt{-g} \left[ \frac{1}{2\kappa} R+ L_{\text{\tiny m}} 
(\varphi, X) \right],
\label{RTaction}
\end{equation}  
where $\kappa$ is a constant, and we have absorbed the infinitesimal $d^{p+1} u$  in the integral sign over $m$, henceforth. To make contact with  general relativity, $\kappa = 8\pi G_N$. The field variables are the embedding functions $X^\mu (u^a)$, and the possible matter fields living on the brane, named collectively 
with $\varphi (u^a)$, with matter Lagrangian $L_{\text{\tiny m}} (\varphi, X)$. 
This action is identical in form to the Einstein-Hilbert action of general relativity, but the field variables
are different. In particular, the induced metric $g_{ab} = X_a \cdot X_b$, that determines the scalar curvature $R$, is a 
composite geometrical structure. We assume that the world volume is without boundary, for simplicity. The symmetries of the action are world volume reparametrizations, the Poincar\'e symmetry of the background Minkowski spacetime and, since $p+1 < n +1$, invariance
under rotations of the normal vectors adapted to the world volume.

The infinitesimal changes of the field variables,
$X^\mu (u^a) \to X^\mu (u^a) + \delta X^\mu 
(u^a)$, can be decomposed into tangential and normal 
deformations as follows $\delta X^\mu = \phi^a (u^b) X^\mu{}_a
+ \phi^i (u^b)  n^\mu{}_i$, where $\phi^a$ and $\phi^i$ denote 
tangential and normal deformations fields, respectively
\citep{capovillageometry}. The tangential deformations can always be
associated to an infinitesimal reparametrization, and can be ignored safely if there is no boundary.
We are left with the physical  transverse deformations, that we denote with
 $\delta_\perp X^\mu = \phi^i n^\mu{}_i$. In particular, we note that
 \begin{equation}
 \delta_\perp g_{ab} = 2 X_{(a} \cdot \delta_\perp X_{b)} =  2 K_{ab}{}^i \phi_i\,,
 \label{eq:deltag}
 \end{equation}
together with $\delta_\perp g^{ab} = - 2 K^{ab}{}_i \phi^i$. 

The first variation of the action (\ref{RTaction}) reads \cite{wald2010general}
\begin{eqnarray}
\delta  S_{\text{\tiny RT}} [X,\varphi]  &=&   
\int_m \left[ \frac{\sqrt{-g}}{2\kappa}  G_{ab}   + 
\frac{ \partial ( \sqrt{-g}  L_{\text{\tiny m}} )}{ \partial g^{ab} } \right]  \delta g^{ab}, 
\n 
\\
&=&   
\int_m \left[ \frac{\sqrt{-g}}{2\kappa}  G_{ab}   + 
\frac{ \partial ( \sqrt{-g}  L_{\text{\tiny m}} )}{ \partial g^{ab} } \right]  \delta_\perp g^{ab}, 
\n 
\end{eqnarray}
where we have neglected a boundary term in the first line, whereas
in the second line we have considered that the complete variation 
of the induced metric, namely, $\delta g_{ab} = 2 K_{ab}{}^i \phi_i 
+ 2 \nabla_{(a} \phi_{b)}$, implies that the tangential deformations
of the world volume give a tangential variation of the action  $\delta S_\parallel$ that is only a boundary 
term. It follows from (\ref{eq:deltag}) that
\be
\delta S_{\text{\tiny RT}} [X,\varphi] =  - \frac{1}{2 \kappa} 
\int_m \sqrt{-g} \left( G_{ab} - \kappa T_{ab} \right) 
K^{ab}{}_i \phi^i\,,
\ee
Here, $G_{ab}$ denotes the world volume Einstein tensor, and 
$T_{ab} := (- 2 /\sqrt{-g}) \partial (\sqrt{-g}
L_{\text{\tiny m}})/\partial g^{ab}$ is the world volume 
stress-energy tensor. The extremization of the action, $\delta S_{\text{\tiny RT}}  = 0$, gives the 
$n-p$ equations of motion
\begin{equation}
\mathcal{E}_i = \left( G_{ab} - \kappa T_{ab}\right)K^{ab}{}_i 
= 0.
\label{eom1}
\end{equation}
These equations are of second-order in  derivatives 
of the fields $X^\mu$, with the extrinsic curvature tensor 
playing the role of an acceleration. This type of gravity has 
a built-in Einstein limit since every solution of Einstein 
equations, $G_{ab} - \kappa T_{ab} = 0$, is necessarily 
a solution of geodetic brane gravity. On the other hand, equations 
(\ref{eom1}) are weaker in the sense that a more general 
solution of the form $G_{ab} - \kappa T_{ab} = \mathcal{T}_{ab}$
may exist as long as
\be
\mathcal{T}_{ab} K^{ab}_i = 0 
\quad \text{and}\quad 
\mathcal{T}_{ab} \neq 0.
\ee
Certainly, it has been speculated 
in \cite{karasik2003geodetic} that the geometrical
structure $\mathcal{T}_{ab}$ can interpreted as a 
non-ordinary matter contribution, labelled as 
\textit{dark matter}, since it is not included in  
the standard matter contribution  $T_{ab}$.

Notice further that, in the absence of matter, in
the spirit of classical string theory, we may think of 
the equations of motion (\ref{eom1}) as the generalization 
of the condition for extremal surfaces in the sense 
that we have the vanishing  of a trace of the 
extrinsic curvature, where the Einstein tensor $G_{ab}$ 
plays the role of the induced metric.

\section{Jacobi equations}
\label{sec4}

Not all solutions of the equations of motion (\ref{eom1}) 
lead to stable configurations of the extended object. In 
this sense, the Jacobi equations provide conditions to 
explore this issue.  Certainly, their solutions, also named  
Jacobi fields, help us to understand more deeply how the 
geometry behaves under deformations of the embedding functions 
in the background spacetime. Since we are interested in 
obtaining a covariant expression for such equations, a 
convenient strategy is to directly  linearize the equations 
of motion (\ref{eom1}).  As formally discussed in \cite{capovillageometry}, for co-dimension higher than one, 
the application of the normal deformation operator 
$\tilde{\delta}_\perp$ to the equations of motion together 
with the assumption that the equations of motion $\mathcal{E}_i 
= 0$ are fulfilled, afford  the linearized equations. Hence, 
we focus on the expressions
\be
\tilde{\delta}_\perp 
\left( G_{ab}K^{ab\,i}-\kappa T^{ab}K_{ab}{}^i\right) = 0.
\label{eq:lin1}
\ee
Here, $\tilde{\delta}_\perp$ denotes a deformation operator
covariant under normal frame rotations, associated with the 
physical transverse motions, and constructed in analogy to 
the covariant derivative $\tilde{\nabla}_a$, 
\citep{capovillageometry}. In reference to this subject, 
it is worth mentioning the existence of another connection, 
$\gamma_{ij} = \gamma_{[ij]}$, analogous to $\omega_a{}^{ij}$,  
that is necessary to satisfy the requirement of the 
covariance of the deformations under normal frame rotations. Fortunately, its incorporation has no physical consequences 
once the on-shell condition is met since $\tilde{\delta}_\perp \mathcal{E}_i = \delta_\perp \mathcal{E}_i - \gamma_i{}^j 
\mathcal{E}_j$. 

The variation $\tilde{\delta}_\perp (G_{ab}K^{ab}{}_i)$
is rather involved. To perform it we will proceed in steps. 
By considering the explicit form of the world volume Einstein 
tensor we have that
\beq
\tilde{\delta}_\perp (G_{ab}K^{ab}_i) &=& \left( \frac{1}{2} 
R K_{ab i} - \frac{1}{2} R_{ab} K_i + 2 G_a{}^cK_{cb i}
\right)\delta_\perp g^{ab} 
\n
\\
&+& G^{ab} \td K_{ab \,i} + K^{ab}{}_i \delta_\perp R_{ab}
\n
\\
&-& \frac{1}{2} K_i g^{ab}\delta_\perp R_{ab}.
\label{var1}
\eeq
Before we continue, we need to compute the variations involving 
the Ricci tensor. From (\ref{gcm3}), after a straightforward
computation we get
\beq
\label{var1-a}
g^{ab} \delta_\perp R_{ab} &=& 2(g^{ab}K_j - K^{ab}{}_j) 
\td K_{ab}{}^j + R_{ab} \delta_\perp g^{ab},
\\
K^{ab}_i \delta_\perp R_{ab} &=& \left( g^{ab} K^{cd}{}_i K_{cd\,j}
-2K^{ac}{}_i K^b{}_{c\,j} 
\right.
\n
\\
&+& \left. K^{ab}{}_i K_j \right)\td K_{ab}{}^j
+ R_{acbd} K^{cd}{}_i \delta_\perp g^{ab}.
\label{var1-b}
\eeq
Inserting these expressions
into (\ref{var1})
yields
\begin{widetext}
\beq
\tilde{\delta}_\perp (G_{ab}K^{ab}{}_i) &=& \left( 
 R_{acbd} K^{cd}{}_i - R_{ab} K_i + \frac{1}{2} R K_{ab\,i}
+ 2 G_a{}^c K_{bc\,i} \right) \delta_\perp g^{ab}
\n
\\
&+& \left[ K_i K^{ab}{}_j + K_j K^{ab}{}_i
- K^a{}_{c\,i} K^{bc}{}_j - K^a{}_{c\,j} K^{bc}{}_i 
- g^{ab} (K_i K_j - K^{cd}{}_i K_{cd\,j}) + G^{ab} 
\delta_{ij} \right] \td K_{ab}{}^j.
\label{var2}
\eeq
\end{widetext}
In turn, this expression suggests to introduce the geometric 
tensor 
\begin{eqnarray}
\mathbb{G}^{ab}{}_{ij}&:=& \left[ K_i K^{ab}{}_j + K_j
K^{ab}{}_i - K^a{}_{c\,i} K^{bc}{}_j - K^a{}_{c\,j} K^{bc}{}_i 
\right.
\n
\\
&-&\left. g^{ab}\left( K_i K_j - K^{cd}{}_i K_{cd\,j}
\right) \right].
\label{Gab}
\end{eqnarray}
Note that it is symmetric in both pairs of indices
$\mathbb{G}^{ab}{}_{ij} = \mathbb{G}^{ba}{}_{ij} 
= \mathbb{G}^{ab}{}_{ji}. $

A fact that will be used below is that this tensor quadratic 
in the extrinsic curvature is divergence-free with respect 
to the tangential indices
\begin{equation}
\wt_a \mathbb{G}^{ab}{}_{ij} = 0.
\label{eq:div}
\end{equation}
This is not self-evident, and it requires the use of the 
contracted Codazzi-Mainardi equations (\ref{whole1}), as 
shown explicitly in Appendix \ref{appA}. 

Note that for a hypersurface, using the contracted Gauss-Codazzi equations (\ref{whole1}), $\mathbb{G}^{abi}{}{ij}$ 
takes the simple form $\mathbb{G}^{ab} = 2 G^{ab}$, a fact 
that emphasizes its geometrical nature, for co=dimension
higher than one.

By inserting (\ref{var2}) into (\ref{eq:lin1}), and 
taking into account  the definition (\ref{Gab}), we get
\begin{widetext}
\be 
\left[ \mathbb{G}^{ab}{}_{ij} + (G^{ab} - \kappa T^{ab})
\delta_{ij} \right] \td K_{ab}{}^j 
+ \left(  R_{acbd} K^{cd}{}_i - R_{ab} K_i + R_a{}^c K_{bc\,i} 
+ G_a{}^c K_{bc\,i} \right) \delta_\perp g^{ab} - \kappa K_{ab\,i}
\delta_\perp T^{ab} = 0.
\label{jacobi1}
\ee
\end{widetext}

At this stage, we are ready to insert the needed normal 
deformations of the first and second fundamental forms
\beq
\delta_\perp g^{ab} &=& - 2 K^{ab}{}_i\,\phi^i,
\label{var3}
\\
\td K_{ab}{}^i &=& - \wt_a \wt_b \phi^i + K_{ac}{}^i K^c{}_{b\,j}
\,\phi^j,
\label{var4}
\eeq
into (\ref{jacobi1}) to obtain
\begin{widetext}
\beq
\left[ \mathbb{G}^{ab}{}_{ij} + (G^{ab} - \kappa T^{ab})
\delta_{ij} \right] \wt_a \wt_b \phi^j &+& \left\lbrace 
2(  R_{acbd} K^{cd}{}_i - R_{ab} K_i + R_a{}^c K_{bc\,i} 
+ G_a{}^c K_{bc\,i})K^{ab}{}_l 
\right.
\n
\\
&& \left. - \left[ \mathbb{G}^{ab}{}_{ij} + (G^{ab} 
- \kappa T^{ab}) \delta_{ij} \right] K_a{}^{c\,j}
K_{bc\,l} \right\rbrace\phi^l + \kappa \,K_{ab\,i}
\delta_\perp T^{ab} = 0.
\label{jacobi2}
\eeq
\end{widetext}
To illustrate the overall  structure of these equations, 
quite formidable in their appearance and content,
it is convenient at this point  to define the tensor
\beq\mathbb{M}^{ab}{}_{ij}:= 
\mathbb{G}^{ab}{}_{ij} + (G^{ab} - \kappa T^{ab}) 
\delta_{ij}\,.
\eeq
By virtue
of this definition, we have the useful identity
\beq
\mathbb{M}^{ab}{}_{ij} K_a{}^{c\,j}K_{bc\,l} &=& 
(R_{acbd} K^{cd}{}_i - R_{ab} K_i + R_a{}^c K_{bc\,i} 
\n
\\
&+& G_a{}^c K_{bc\,i})K^{ab}{}_l - \kappa\,T_a{}^cK_{bc\,i} 
K^{ab}{}_l,
\n
\eeq
that coincides with the terms appearing on the r.h.s. of 
(\ref{jacobi2}). This identity allows us to rewrite  
(\ref{jacobi2}) in a compact form as 
\beq
\mathbb{M}^{ab}{}_{ij} \wt_a \wt_b \phi^j &+& \left( 
\mathbb{M}^{ab}{}_{il}K_a{}^{c\,l}K_{bc\,j} 
+ 2\kappa T_a{}^c K_{bc\,i} K^{ab}{}_j \right) \phi^j 
\n
\\
&+& \kappa\,K_{ab\,i} \delta_\perp T^{ab} = 0.
\label{jacobi3}
\eeq
These are the Jacobi equations for geodetic brane gravity 
describing small normal deformations of the world volume.
As a matter of course, the inclusion of arbitrary matter 
fields confined to the world volume, affects the dynamics 
of the Jacobi fields by including the derivatives of the 
matter fields. Note that (\ref{jacobi3}) are second-order 
partial differential equations in the unknown functions  
$\phi^i$, which is a feature that characterizes brane 
theories with second-order derivative equations of motion \citep{bagatella2016covariant}. As mentioned earlier, the  
solutions to the Jacobi equations address the issue of 
stability through the nature of the normal modes $\phi^i$, 
so appropriate boundary conditions must enter the game.

On the other hand, if we focus our attention on the case 
where there are  no  brane matter fields, $T_{ab} = 0$, and 
assume the fulfillment of the equations of motion,
we obtain a more compact and elegant expression for the
Jacobi equations for a pure RT geometrical model. Indeed, by 
defining now a new tensor
\beq
\mathcal{M}^{ab}{}_{ij}
:= \mathbb{G}^{ab}{}_{ij} + G^{ab}\delta_{ij},
\eeq 
we can write (\ref{jacobi3}) in the form
\be 
\mathcal{M}^{ab}{}_{ij} \wt_a \wt_b \phi^j + 
\mathcal{V}_{ij}
\,\phi^j = 0,
\label{jacobi4}
\ee
where we identify a geometrical  ``potential"
\be
\mathcal{V}_{ij} := \mathcal{M}^{ab}{}_{il}
K_a{}^{c\,l} K_{bc\,j}.
\label{Vij}
\ee
The formal similarity of (\ref{jacobi4}) with a set of 
Klein-Gordon equations is indeed striking. In passing, 
note that the matrix structure (\ref{Vij}) is symmetric 
in the normal indices.

This set up paves the way to construct an accessory variational problem. Under these conditions, observe that the kinetic 
``mass matrix" $\mathcal{M}^{ab}{}_{ij}$ is divergenceless, 
as follows from the geometric identity (\ref{eq:div}), and the 
divergenceless property of the Einstein tensor,
\beq
\wt_a  \mathcal{M}^{ab}{}_{ij} = 0.
\eeq

\sk
The accessory action can be written then as
\begin{equation}
S_{\text{\tiny RT}}  [ \phi ] = - \frac{1}{2} \int_m 
\sqrt{-g} \left[ \mathcal{M}^{ab}{}_{ij} \wt_a \phi^i 
\wt_b \phi^j - \mathcal{V}_{ij} \phi^i \phi^j 
\right].
\label{acc}
\end{equation} 
Up to a boundary term, variation with respect to the 
normal deformations gives the Jacobi equations in the 
form (\ref{jacobi4}) as its Euler-Lagrange equations. It 
is interesting to note that the accessory principle, up to 
factor of one half, gives the {\it index} of the RT 
geometric model,
\begin{equation}
I_{\text{\tiny RT}}  [\phi] = \int_m {\mathcal I} (\phi,  
\phi)\,.
\end{equation} 

As all accessory variational principles, (\ref{acc}) is quadratic 
in the field variables. This makes it suitable for a 
quantization using a path integral approach, and a 
determination of the effect of quantum fluctuations 
\cite{zinn2021quantum}. We plan to address this issue
in future  work.

Regarding the hypersurface case, by considering the 
reductions $\mathbb{G}^{ab} = 2 G^{ab}$ and $\mathcal{M}^{ab}
= 3 G^{ab}$, the Jacobi equations (\ref{jacobi4})
specialize to
\begin{equation}
G^{ab}\left( \nabla_a \nabla_b\phi+K_a{}^cK_{bc}\phi
\right)=0.
\label{eqcodimension1}
\end{equation}
This result is in agreement with the one found in \citep{bagatella2016covariant}, where the RT model is 
considered as a special case of Lovelock branes.

Clearly, the Jacobi equation \eqref{eqcodimension1} 
can be obtained from the extremization of the action 
functional  
\begin{equation} 
S [\phi] = - \frac{1}{2}\int_m \sqrt{-g}\left[ G^{ab}\nabla_a\phi \nabla_b\phi - G^{ab}K_a{}^cK_{bc} \,\phi^2  \right],
\end{equation}
when  varied with respect to the $\phi$ field.
If we consider a world volume such that $G_{ab} 
\sim g_{ab}$, {\it i.e.} an Einstein manifold, 
this action reduces to the action of a massive scalar 
field where the variable mass term is proportional 
to $K_{ab}K^{ab}$. This result is equivalent to what 
is obtained when one linearizes the equation of the 
DNG model in a  flat background, see \cite{guven1993covariant}.

Another interesting case is provided by the inclusion
of the DNG action, playing the role
of a cosmological constant $\Lambda$, in our development.
In such a case, $L_{\text{\tiny{m}}} = \Lambda$ so that
$T_{ab} = \Lambda \,g_{ab}$. The form of the Jacobi
equations, (\ref{jacobi4}), remains unchanged except
that the matrix $\mathbb{M}^{ab}{}_{ij}$ now becomes
\be 
\mathbb{M}^{ab}{}_{ij} = \mathbb{G}^{ab}{}_{ij}
+ (G^{ab} - \kappa \Lambda\,g^{ab})\delta_{ij}.
\ee
Notice that we still have at hand the divergenceless 
property $\wt_a \mathbb{M}^{ab}{}_{ij} = 0$.

\section{Linear stability of Schwarzschild geometry in $\mathcal{M}^6$}
\label{sec5}

To appreciate the formalism previously developed, we consider 
the special case of a Schwarzschild geometry for the world volume, 
with no brane matter fields, embedded in a 6-dim Minkowski 
spacetime, $\mathcal{M}^6$. A 
Schwarzschild solution of general relativity is also 
automatically a particular solution of the equations of 
motion (\ref{eom1}). We  use the Jacobi equations to study 
its linear local stability. An illustrative case, with still 
a high degree of complexity in the search for its analytical 
solution, is provided by the particular setup $G_{ab} = R_{ab}
= T_{ab} = 0$. Under these assumptions \eqref{jacobi4} become
\be
\mathbb{G}^{abij}\left(\wt_a\wt_b\phi_j 
+ K_{ac\,j}K^c{}_{b\,l}\,\phi^l \right)= 0.
\label{lnbh}
\ee

Among the different embeddings for a 4-dim  
Schwarzschild geometry, see {\it e.g.} 
\cite{paston2012embeddings}, we choose to 
consider the Fronsdal embedding  \cite{fronsdal1959completion,Davidson2000} given by
\begin{equation}
\begin{split}
&X^1 = 2R \sqrt{1 - \frac{R}{r}} \sinh \left( \frac{t}{2R} 
\right),
\\
&X^2 = 2R \sqrt{1 - \frac{R}{r}} \cosh \left( \frac{t}{2R} 
\right),
\\
&X^3 = \int \sqrt{\frac{R}{r} 
+ \left( \frac{R}{r}\right)^2
+ \left( \frac{R}{r}\right)^3}dr,
\\
&X^4=r \sin\theta\sin\phi,\\
&X^5=r \sin\theta\cos\phi,\\
&X^6=r \cos\theta
\end{split}
\end{equation}
where $\lbrace t,r,\theta,\phi\rbrace$ are the local brane 
coordinates, and $R$ denotes the event horizon. For this 
parametrization, we have two non-vanishing extrinsic curvature components  given by
\begin{equation}
\begin{split}
&K_{ab}{}^1=\text{diag}\left( 0, ba'-ab',-rb,-rb\sin^2
\theta\right),
\\
&K_{ab}{}^2=\text{diag}\left( - \frac{\gamma^2}{2Ra}, 
\frac{2aR^2}{r^3 - Rr^2}, - \frac{aR^2}{r}, 
- \frac{aR^2}{r \sin^2 \theta} \right),
\end{split}
\end{equation}
where we have introduced
\begin{equation}
\begin{split}
&a=\sqrt{\frac{r^3}{r^3 + r^2 R + r R^2 + R^3}},\\
&b=\sqrt{\frac{R (r^2 + r R + R^2)}{r^3 + r^2 R + r R^2 + R^3}},\\
&\gamma =\sqrt{1-\frac{R}{r}}.
\end{split}
\end{equation}

By considering the ansatz  $\Phi=e^{-i\omega t} Y_{lm}
(\theta,\phi) \rho(r)$, where the normal deformation 
field is $\Phi=(\phi^1,\phi^2)$,  and $\rho$ is also 
be considered as a vector field, $\rho = (\rho^1,\rho^2)$, 
the Jacobi equations \eqref{jacobi4} are 
separable, and can be written as a matrix arrangement of radial equations
in the form
\be
{\bf A}\rho''+{\bf B}\rho'+({\bf C}-\omega^2 {\bf D})
\rho = 0,
\label{explieq1}
\ee
where $\rho' = d\rho / dr$. Here, ${\bf A}$, ${\bf B}$, 
${\bf C}$, and ${\bf D}$ are $2\times 2$ matrices  that 
depend on the radial coordinate $r$. Their explicit 
components are given in Appendix \ref{appB}. It is worth 
mentioning that the matrix ${\bf C}$ is the only one 
that contains the angular momentum information through 
$l$. Multiplying  equation \eqref{explieq1} by ${\bf D}^{-1}$, and introducing the tortoise-like radial coordinate 
$r_*=\int (dr/f_g)$ with $f_g=br^2\gamma^2 / \sqrt{3}aR^2$, 
and considering $\rho= \mathbf{M} \chi$, where $\mathbf{M}$ 
is a matrix defined in such a way that the term proportional to 
$d \chi/ dr_*$ vanishes, then the system of equations 
\eqref{explieq1} acquires a form familiar  in black hole
theory stability analysis
\begin{equation}
\frac{d^2\chi}{dr_*^2}+\omega^2\chi-\mathbf{V} \chi=0,
\label{oscieq}
\end{equation} 
where, as  in (\ref{explieq1}), $\chi$ 
must be understood as a vector. The  matrix potential 
$\mathbf{V}$ is explicitly written in terms of matrices 
${\bf B,\ C,}$ and, ${\bf D}$ in Appendix \ref{appB}. 
The system of equations \eqref{oscieq} provides a system 
of coupled harmonic oscillators with quasi-normal 
frequencies $\omega=\omega_R+i\omega_I$. One can test the 
stability of this configuration by studying these frequencies 
$\omega$. 

\begin{figure}[htb!]
\centering
\includegraphics[width=85mm]{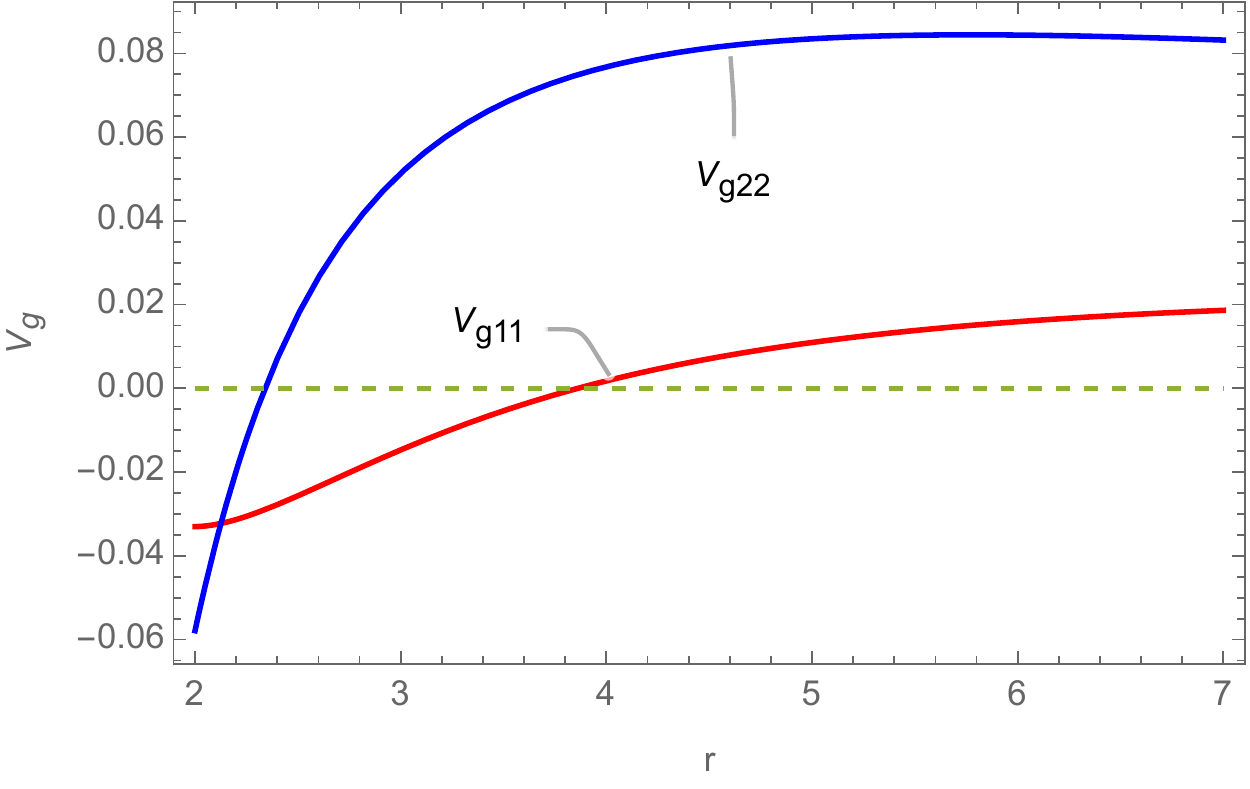}
\caption{$V_{g11}(r)$ and $V_{g22}(r)$  for $l=1$ and $R=2$.}
\label{fig1}
\end{figure}

Regarding the asymptotic behavior of the fields, at spatial 
infinity for non zero angular momentum, $l \neq 0$, the 
values of the diagonal components of the potential matrix 
$\mathbf{V}$ diverge. It follows that one can assume that 
the field $\chi$ must be zero at $r \to \infty$.  On the 
other hand, the matrix potential $\mathbf{V}$ vanishes at 
the event horizon $R$, so the solution to the resulting 
equation is $\chi\sim e^{-i\omega r_*}+ e^{i \omega r_*}$, 
where the exponential with sign - (+) represents an incoming 
(outgoing) wave to the black hole configuration. Additionally, 
since nothing can escape from the black hole, the part of 
the solution corresponding to an outgoing wave is not allowed. 
Thus, the field $\chi$ must be of the form $\chi = 
e^{-i \omega r_{*}} \psi$. Substituting this in \eqref{oscieq}, 
the following set of equations is obtained
\begin{equation}
f_g\frac{d}{dr}\left( f_g \frac{d \psi}{d r}\right)-2 i 
\omega f_g \frac{d\psi}{d r}-\mathbf{V}\psi=0.
\label{subsol}
\end{equation}
We divide this equation by $f_g$, and then multiply it by 
$\psi^{\dag}$ so that we get 
\begin{equation}
\psi^{\dag}\frac{d}{dr}\left( f_g\frac{d\psi}{dr}\right)-2 
i\omega \psi^{\dag}\frac{d\psi}{dr}-\psi^{\dag}\mathbf{V}_g
\psi=0,
\label{eqmul}
\end{equation}
where $\mathbf{V}_g=\mathbf{V}/f_g$. Now, integrating by 
parts \eqref{eqmul}, and taking into account that $\psi(\infty)
=0$ and $f_g(R) =0$, we obtain
\begin{equation}
\int_R^{\infty}dr\left[ f_g\left\vert \frac{d\psi}{dr}
\right\vert^2 +2 i\omega \psi^{\dag}\frac{d\psi}{dr}
+ \psi^{\dag}\mathbf{V}_g\psi \right] = 0.
\label{eqmul2}
\end{equation}
The transpose and conjugate operations applied to the last 
equation result in
\begin{equation}
\int_R^{\infty}dr \left[ f_g\left\vert \frac{d\psi}{dr}
\right\vert^2 - 2i\omega^{*}\frac{d\psi^{\dag}}{dr} \psi 
+ \psi^{\dag}\mathbf{V}_g^{\dag}\psi \right]=0.
\label{eqmu3}
\end{equation}
We proceed to integrate by parts the second term of the 
previous equation, and then we take the difference 
of the result with \eqref{eqmul2}. We get
\beq
&&\int_R^{\infty}dr \left[ (\omega-\omega^*)\psi^{\dag} \psi' - \frac{i}{2} \psi^{\dag}\left( \mathbf{V}_g-\mathbf{V}_g^{\dag}
\right)\psi\right] 
\n
\\
&& \qquad \qquad \qquad  \qquad \qquad \qquad \qquad \quad
 = \omega^{*}\left\vert\psi (R)\right\vert^2,
\label{sut}
\eeq
where $\psi' = d\psi/dr$. From \eqref{sut} we can solve 
for $\psi^{\dag}\psi'$ and substituting this in the 
\eqref{eqmul2}, we obtain 
\begin{widetext}
\begin{equation}
\int^{\infty}_Rdr\left[ f_g\left\vert \psi' \right\vert^2
+\frac{i\omega_R}{2\omega_I}\psi^{\dag}\left( \mathbf{V}_g
-\mathbf{V}_g^{\dag}\right)\psi+\frac{1}{2} \psi^{\dag}\left( \mathbf{V}_g+\mathbf{V}_g^{\dag}\right)\psi\right]
=- \frac{\left\vert\omega\right\vert^2\left\vert \psi (R)
\right\vert^2}{\omega_I},
\label{eqstb1}
\end{equation}
\end{widetext}
where we assume that $\omega_I \neq 0$. 
If the integrand in \eqref{eqstb1} is positive 
definite,
then the imaginary part of frequency must be negative, that 
is an indication of having stable deformations of this black 
hole geometry. Indeed, since the first term in \eqref{eqstb1} 
is positive, it only remains to analyse the nature of the 
second term
\begin{widetext}
\begin{equation}
\begin{split}
\nu :=&\frac{i\omega_R}{2\omega_I}\psi^{\dag}\left( \mathbf{V}_g-\mathbf{V}_g^{\dag}\right)\psi+\frac{1}{2} \psi^{\dag}\left( \mathbf{V}_g+\mathbf{V}_g^{\dag}\right)\psi
\\
=&V_{g11}\left\vert\psi_1\right\vert^2+V_{g22}\left\vert\psi_2\right\vert^2-\left\vert \left( V_{g12}+V_{g21}\right)+\frac{\omega_R}{\omega_I}\left( V_{g12}-V_{g21}\right)\right\vert \vert\psi_1\psi_2\vert\\ 
&+\left( \left\vert\psi_1\right\vert^2  +\left\vert\frac{1}{2}\left( V_{g12}+V_{g21}\right)\psi_2\right\vert^2 +\left\vert V_{g12}+V_{g21}\right\vert \vert \psi_1\psi_2 \vert -\left\vert \psi_1-\frac{1}{2}\left( V_{g12}+V_{g21}\right)\psi_2\right\vert^2\right) \\
&+\left( \left\vert\psi_1\right\vert^2  +\left\vert\frac{\omega_R}{2\omega_I}\left( V_{g12}-V_{g21}\right)\psi_2\right\vert^2 +\left\vert \frac{\omega_R}{\omega_I}\right\vert \vert V_{g12}-V_{g21}\vert \vert \psi_1\psi_2\vert-\left\vert \psi_1-\frac{i\omega_R}{2\omega_I}\left( V_{g12}-V_{g21}\right)\psi_2\right\vert^2\right),
\end{split}
\end{equation}
\end{widetext}
where a matrix multiplication has been performed in the 
second line, and we have used the triangle inequality. Note 
that the two terms that appear in the parentheses are 
positive definite. Therefore, the sign of $\nu$ depends 
strongly on the values of $V_{g11}$ and $V_{g22}$. 
In this sense, $\nu$ could be negative if $V_{g11}$ and 
$V_{g22}$ are both negative enough. Indeed, to illustrate
this fact, in Figure \ref{fig1} we show the functions 
$V_{g11}$ and $V_{g22}$ for $l=1$ and $R=2$.
This fact is an indication that, for $r > 2$, there are 
negative values of $V_{g11}$ and $V_{g22}$, so there could 
exist frequencies with positive imaginary part that are 
associated with an unstable oscillation mode. However, this 
analysis is not conclusive since the value of the integral 
in \eqref{eqstb1} could still be positive. 

A suitable strategy to further the analysis of the 
frequencies is to apply  numerical methods. 
Fortunately, a variety of numerical methods have been 
considered in order to solve this type of problems, such 
as continued fraction, series method, \cite{ferrari2007new,pani2012perturbations,horowitz2000quasinormal,pani2013advanced}, and so on, and one has the option to choose the more suitable  
one according to the asymptotic behavior of the potential 
both at spatial infinity and at the event horizon.
The matrix potential ${\bf V}$, at the boundaries, behaves 
as the potential of Schwarzschild black hole in anti-de Sitter spacetime evaluated in the same regions. In this sense, we rely 
on the numerical method carried out in \cite{horowitz2000quasinormal} for guidance in our case, where a similar analysis is performed. 
By considering a Taylor expansion of the components of field 
$\chi$ as follows
\begin{equation}
\chi_ i=e^{-i\omega r_*}\sum_n a^{(i)}_n(\omega)\left(r-R\right)^n,
\label{fieldexpan}
\end{equation}
we can substitute this into \eqref{oscieq}, and expand those differential equations in a Taylor series around the event 
horizon. Then, by solving the resulting algebraic equations, 
order by order, the coefficients $a^{(i)}_n(\omega)$ can be 
found. We observe that they only depend on the 2-dimensional 
vector $\mathbf{a_0}=\lbrace a_0^{(1)},a_0^{(2)}\rbrace$. In 
this sense, we can choose a suitable orthogonal basis for 
the 2-dimensional space of the initial coefficients $a_0^{(1)}$ 
and $a_0^{(2)}$. For each element of the basis, one constructs 
the $2\times 2$ matrix
\begin{equation}
\mathbf{S}(\omega) = \lim_{r\rightarrow \infty}\left(
\begin{matrix}
\chi_1^{(1)}&\chi_1^{(2)}
\\
\chi_2^{(1)}&\chi_2^{(2)}
\end{matrix}
\right),
\end{equation}
where  the superscripts denote a particular vector of the 
basis. Certainly, $\chi_{i}^{(1)}$ corresponds to $\mathbf{a_0}
=\lbrace 1,0\rbrace$ and $\chi_i^{(2)}$ corresponds to 
$\mathbf{a_0}=\lbrace 0,1\rbrace$ (see \citep{pani2013advanced} 
for more specific details in the procedure). Following the 
line of reasoning in \citep{pani2013advanced}, to obtain the 
values of the frequencies $\omega$ it is necessary to find 
the zeros of (\ref{fieldexpan}) by imposing the condition
\be
\det \mathbf{S}(\omega)=0.
\label{detS}
\ee
As mentioned previously, the process indicated in (\ref{detS}) 
can be coded in \textsc{Mathematica}, and one can find the 
lowest eigenfrequencies for different values of $l$ and $R$. 
Indeed, taking the following variable change $r\rightarrow Rr$ 
in \eqref{oscieq}, we are able to find the values shown in 
Table~\ref{table:modes}.
\begin{table}[htbp]
\center
\begin{ruledtabular}
\begin{tabular}{lll}
$l$ & $R\,\omega_R$ & $R\,\omega_I$\\
\hline 
1 & $\sim$ 0 & 0.83 \\ 
2 & 1.01 & 0.36 \\
3 & 1.62 & 0.56 \\ 
4 & 2.45 & 0.75 \\ 
\end{tabular}
\end{ruledtabular}
\caption{Frequencies for different values of $l$.}
\label{table:modes}
\end{table}
The  value $\omega_R$ is associated with the oscillation of the deformation while  $\omega_I$ is related to  its attenuation or increasing.
Note that the imaginary part of $ \omega $  is always positive  for any value of $ R $. These frequencies are associated with unstable deformations since the  field is $\Phi\sim e^{-i \omega t}$, and this diverges when $t\rightarrow \infty$. 
Although a Schwarzschild four-dimensional black hole is stable in general relativity, our results show that linear instabilities are present in the embedded Schwarzschild black hole that satisfies the geodetic brane gravity equation. This kind of instabilities is also found in the study of higher-dimensional black holes, see {\it e.g.} 
 \cite{gregory1993black}. 

\section{Discussion}
\label{sec6}

In this paper we have developed a covariant approach for the 
analysis of perturbations of branes  governed 
by  geodetic brane gravity through the Regge-Teitelboim geometric model, in any co-dimension.

We focus on normal deformations since they are the only ones that represent 
the physical perturbations. We have obtained the covariant  Jacobi equations for the model, that determine
the behaviour of such breathing perturbations modes $\phi^i$. In particular, the 
equations are explicitly covariant under local normal frame 
rotations. Within this geometric framework, some conserved 
geometric structures enter the game, and help to express the 
Jacobi equations as a set of wave-like forms from which 
geometric ``mass'' terms can be read off.  There is consistency 
in the theory since by making the appropriate reduction for 
co-dimension one, results previously found are met \cite{bagatella2016covariant}.

In addition, we were able to exploit the conserved geometrical structures to construct an action, or Jacobi accessory variational
principle, whose extremalization produces the Jacobi equations as   its equations  of motion. This action is proportional to the geometric index for the RT model. It is worth emphasizing that it is quadratic
in the field modes, opening an interesting avenue in the path integral quantization of the model,
that we plan to investigate in future work.

As an illustration of the general framework,  using a 
Fronsdal scheme for a black-hole geometry, our general 
result has been specialized to a Schwarzschild geometry 
embedded in $\mathcal{M}^6$. A study of the Jacobi 
equations for this case using analytical methods is 
instructive, but not conclusive. On the other hand, 
supported by a numerical analysis, unstable small 
deformations are found. In this sense, it should be emphasized 
that our discussion of the properties of the eigenmodes $\omega$, related to the stability of this type of black holes 
embedded in a higher dimensional space, is neither exhaustive 
nor complete, but we believe that our approach opens the 
door to exploring the stability of interesting relativistic 
systems from covariant expressions obtained in the framework 
of extended objects. Additionally, the results presented 
indicate that they can be generalized to the whole class of 
Lovelock branes, and we consider it as a next step in the 
understanding of this type of geometrical models for branes.

One crucial assumption, both in physical and geometrical 
terms, is the embedding in a flat background spacetime of 
the world volume. A generalization to arbitrary background 
spacetimes is difficult, but for maximally 
symmetric ambient spacetimes like a de Sitter or 
anti-de Sitter background it appears to be doable. We plan to 
address this issue in a future communication.

\section*{Acknowledgements}

GC acknowledges support from a CONACYT-M\'exico 
doctoral fellowship. ER acknowledges encouragement 
from ProDeP-M\'exico, CA-UV-320: \'Algebra, Geometr\'\i a 
y Gravitaci\'on.  Also, RC and ER thank partial support 
from Sistema Nacional de Investigadores, M\'exico.

\appendix
\section{Proof for the conservation law (\ref{eq:div}).}
\label{appA}

In this appendix, we show explicitly that the tensor $\mathbb{G}_{ab}{}^{ij} $ is divergenceless.
We choose to define the tensorial matrices 
\be
\mathbb{G}_{ab}{}^{ij} :=  2 \mathbb{R}_{ab}{}^{ij} - g_{ab} \mathbb{R}^{ij},
\label{eq1}
\ee
where
\beq
\mathbb{R}_{ab}{}^{ij} &:=&  K^{(i} K_{ab}{}^{j)} -  K_a{}^{c(i} K_{bc}{}^{j)} \\
\mathbb{R}^{ij} &:=& K^i K^j - K^{cd\,i} K_{cd}{}^j.
\eeq

The divergence of (\ref{eq1}) is 
\be
\wt^a \mathbb{G}_{ab}^{ij} = 2 \wt^a \mathbb{R}_{ab}^{ij} - g_{ab} \wt^a \mathbb{R}^{ij}.
\n
\ee

On the one hand, we have for the first term
\beq 
 \wt^a R_{ab}^{ij} &=&  (\wt^a K^{(i} ) \, K_{ab}{}^{j)} + (\wt^a K_{ab}{}^{(i} ) \, K^{j)} \n \\
&-& (\wt^a K_a{}^{c\,(i} ) \,
K_{bc}{}^{j)} -
  K_a{}^{c\,(i} \,\wt^a K_{bc}^{j)} 
\n
\\
&=&  (\wt^a K^{(i} ) \, K_{ab}{}^{j)} +(\wt_b K^{(i} ) \, K^{j)} \n\\
&-& ( \wt^c K^{(\,i} )\,K_{bc}^{j)} 
- K_a{}^{c\,(i}\, \wt^a K_{bc}^{j)}
\n
\\
&=& 
(\wt_b K^{(i} ) \, K^{j)} - K_a{}^{c\,(i}\, \wt^a K_{bc}^{j)}
\label{eq:a4}
\eeq
where we have used the contracted integrability condition 
$\wt^a K_{ab}^i = \wt_b K^i$ (Codazzi-Mainardi) to obtain the second line.

On the other hand, observe that for the second term
\beq
\wt_b \mathbb{R}^{ij} = 2 \left[  (\wt_b K^{(i} ) \, K^{j)} - K_a{}^{c\,(i}\, \wt^a K_{bc}^{j)} \right] \,.
\label{eq:a5}
\eeq

The difference between twice (\ref{eq:a4}) and (\ref{eq:a5}) reads
\beq
2 \wt^a R_{ab}^{ij} - g_{ab} \wt^a R^{ij} =  \wt^a G_{ab}^{ij} = 0\,.
\eeq

\section{Explicit form of transformation matrix ${\bf M}$ and effective potential ${\bf V}$.}
\label{appB}
We provide here the explicit form of the 
matrices appearing in \eqref{explieq1}, and write the matrices 
${\bf M}$ and ${\bf V}$ in terms of them. We have
\\
\\
\\
\\
\\
\begin{equation}
\begin{split}
&\mathbf{A}=\left(
\begin{matrix}
a_{11}&a_{12}
\\
a_{12}&-a_{11}
\end{matrix}
\right),\ \ 
\mathbf{B}=\left(
\begin{matrix}
b_{11}&b_{12}
\\
b_{12}&-b_{11}
\end{matrix}
\right),\\
&\mathbf{C}=\left(
\begin{matrix}
c_{11}&c_{12}
\\
c_{12}&-c_{11}-\frac{6R^2}{r^6}
\end{matrix}
\right),\ \ 
\mathbf{D}=\left(
\begin{matrix}
d_{11}&d_{12}
\\
d_{12}&-d_{11}
\end{matrix}
\right).
\end{split}
\end{equation}
Notice that  ${\bf A}$, ${\bf B}$ and ${\bf D}$ are 
traceless symmetric matrices while ${\bf C}$ is not.
In fact, this is  responsible for not being able to
decouple the system of equations \eqref{explieq1}. The explicit components of these matrices are
\begin{widetext}
{\small
\begin{equation}
\begin{split}
&a_{11}=\frac{2 R \left(r^3-R^3\right)}{r^3 \left(r^3+r^2 R+r R^2+R^3\right)},
\\
&a_{12}=-\frac{(r-R)(r^3+r^2R+rR^2-R^3)\sqrt{r^3R(r^2+rR+R^2)}}{r^5R(r^3+r^2R+rR^2+R^3)},
\\
&b_{11}=\frac{2 R \left(-r^6+r^4 R^2+6 r^3 R^3+3 r^2 R^4+2 r R^5+R^6\right)}{r^4 \left(r^3+r^2 R+r R^2+R^3\right)^2},
\\
&b_{12}=\frac{-r^9-3 r^8 R-6 r^7 R^2-22 r^6 R^3-20 r^5 R^4
-16 r^4 R^5+6 r^3 R^6+6 r^2 R^7+5 r R^8+3 R^9}{2 r^3 \sqrt{r^3 R \left(r^2+r R+R^2\right)} \left(r^3+r^2 R+r R^2+R^3\right)^2},
\\
&c_{11}=l(l+1)\frac{R (r-R) \left(r^2+2 r R+3 R^2\right)}{r^5 \left(r^3+r^2 R+r R^2+R^3\right)}-\frac{R^2 \left(9 r^9+27 r^8 R
+54 r^7 R^2+42 r^6 R^3+17 r^5 R^4-21 r^4 R^5+4 r R^8
+12 R^9\right)}{2 r^6 \left(r^3+r^2 R+r R^2+R^3\right)^3},
\\
&c_{12}=l(l+1)\frac{\sqrt{\frac{r^3 R}{r^2+r R+R^2}} 
\left(r^6+2 r^5 R+3 r^4 R^2+16 r^3 R^3+15 r^2 R^4+14 r R^5-3 R^6\right)}{4 r^7 R \left(r^3+r^2 R+r R^2+R^3\right)}
\\
&-\frac{R \left(3 r^{11}+12 r^{10} R+30 r^9 R^2+90 r^8 R^3+182 r^7 R^4+296 r^6 R^5+254 r^5 R^6+150 r^4 R^7-5 r^3 R^8-56 r^2 R^9-56 r R^{10}-36 R^{11}\right)}{4 r^{11/2} \sqrt{R \left(r^2+r R+R^2
\right)} \left(r^3+r^2 R+r R^2+R^3\right)^3},
\\
&d_{11}=-\frac{6 R^4}{r^6-r^2 R^4},
\\
&d_{12}=\frac{3 R^2 \left(r^3+r^2 R+r R^2-R^3\right) \sqrt{r^3 R \left(r^2+r R+R^2\right)}}{(r^7-r^4 R^3)\left(r^3+r^2 R+r R^2+R^3\right)}.
 \end{split}
\end{equation}}
\end{widetext}
On account of the definition
\begin{equation}
{\bf h}:= -\left(\frac{{\bf D}^{-1}{\bf B}}{f_g}
+{\bf I}\frac{df_g}{dr}\right),
\end{equation}
where ${\bf I}$ denotes the $2\times 2$ unit matrix, 
we find that the matrix $\mathbf{M}$ can be 
written as 
\begin{equation}
\mathbf{M}=\exp\left( -\frac{1}{2}\int 
\frac{{\bf h}}{f_g}dr\right).
\end{equation}
In the same way, the potential matrix 
${\bf V}$ in terms of these matrices, becomes
\begin{equation}
{\bf V}=\frac{f_g}{2}\frac{d{\bf h}}{dr}
+\frac{1}{4}{\bf h}^2+{\bf M}^{-1}{\bf D}^{-1}{\bf C} {\bf M}.
\end{equation}

\bibliography{bibpaper}

\end{document}